\documentclass{article}
\usepackage{spconf,amsmath,graphicx,booktabs,multirow}

\title{Soft-Weighted CrossEntropy Loss for Continous Alzheimer's Disease Detection}
\name{Xiaohui Zhang, Wenjie Fu, Mangui Liang}
\address{Department of Computer Science,\\ Beijing JiaoTong University, Beijing 100044, China}

\address{Department of Computer Science,\\ Beijing JiaoTong University, Beijing 100044, China}
\twoauthors
 {Xiaohui Zhang, Mangui Liang}
	{Department of Computer Science \\
	Beijing JiaoTong University \\
	Beijing 100044, China}
 {Wenjie Fu}
	{Research Center of 6G Mobile Communications, \\
 School of Cyber Science and Engineering, \\ 
 and Wuhan National Laboratory for Optoelectronics \\
	Huazhong University of Science and Technology \\
	Wuhan 430074, China}
\begin{document}
%
\maketitle
\begin{abstract}
Alzheimer's disease is a common cognitive disorder in the elderly. Early and accurate diagnosis of Alzheimer's disease (AD) has a major impact on the progress of research on dementia. At present, researchers have used machine learning methods to detect Alzheimer's disease from the speech of participants. However, the recognition accuracy of current methods is unsatisfactory, and most of them focus on using low-dimensional handcrafted features to extract relevant information from audios. This paper proposes an Alzheimer's disease detection system based on the pre-trained framework Wav2vec 2.0 (Wav2vec2). In addition, by replacing the loss function with the Soft-Weighted CrossEntropy loss function, we achieved 85.45\% recognition accuracy on the same test dataset.
\end{abstract}
\section{Introduction}
\label{sec:intro}
Alzheimer's disease (AD) is the most common dementia, accounting for 50\% to 70\% of all types of dementia \cite{2021Alzheimer}. The diagnostic procedure of Alzheimer's disease requires comprehensive examination by medical experts, which requires a lot of cost and time. At present, researchers try to use machine learning-based methods to detect Alzheimer's disease continuously \cite{qiao2021alzheimer, zhang2023remember, zhang23DoYou}, Regularized Adaptive Weight Modification and Radian Weight Modification, shows great performance in this field, or multimodal learning \cite{zhang2023multimodal}. Research shows that patients not only show language impairment in the early stage of AD, but also their language performance in youth can predict the disease in the elderly. Therefore, language based automatic detection and screening method is an important field in recent years.\\
At present, speech based detection of Alzheimer's disease mainly uses handcrafted acoustic features, including Mel-frequency cepstral coefcients (MFCC), the fundamental frequency (F0) semitone, loudness, spectral flux, F1, F2, F3 etc. These handcrafted features are considered to have the ability to detect psychological changes in speech. Saturnino el et \cite{luz2021detecting} extracted the 88 dimensional acoustic features (eGeMAPS) \cite{eyben2015geneva} per 100ms in the INTERSPEECH2021 Alzheimer's Dementia Recognition through Spontaneous Speech Challenge dataset, and used K Nearest Neighbour (KNN), linear discriminant analysis (LDA), Tree Bagger (TB) and support vector machines (SVM) to detect Alzheimer's disease from speech data.\\
With the development of Alzheimer's disease detection technology based on machine learning, researchers begin to explore more feature patterns to improve the performance of recognition system. Syllable tokens were choice from phonemes, syllables, words and utterances by Yi el et \cite{chien2019automatic}, where each token was responsible for representing one unique element in the speech and contained much information about the spoken content. They also add a silent token to the token space and set a threshold for judging whether a silence segment should be transcribed to a silence token based on its length. The Convolutional Recurrent Neural Network (CRNN) trained by Connectionist Temporal Classification (CTC) loss and bidirectional Recurrent Neural Network (RNN) were chosen as the feature sequence generator and the downstream model respectively. \\
In this paper, we introduce an Alzheimer's disease detection system based on a pre-trained model and a downstream model. We apply the pre-trained model XLSR-53 \cite{conneau2020unsupervised} as the feature generator based on the Wav2vec2 framework \cite{baevski2020wav2vec}. This model uses Multilingual LibriSpeech, CommonVoice and Babel about 5,6000 hours of speech data for multilingual self-supervised pre-training, and has achieved excellent performance in the field of Automatic Speech Recognition (ASR) \cite{xu2021self}. In order to make the features extracted by XLSR-53 preserve abundant information conducive to the detection of Alzheimer's disease, we select the speech data of NCMMSC2021 Alzheimer's disease recognition challenge dataset to self-supervised fine-tune the pre-trained model. In addition, we study the joint learning strategy of feature generator and downstream classifier. The experimental results indicate that the parameters of generator need to be frozen in the first N epochs of joint learning. We present the N value that makes the classifier achieve supreme recognition accuracy on our experiments. Finally, in order to further improve the accuracy of Alzheimer's disease recognition system, we designed a loss function called Soft-Weighted Cross Entropy (SWCE) loss. The purpose is to make the model predict more accurate results by improving the recognition probability of ground truth. Through experiments, we compare the effects of SWCE loss and cross entropy loss on recognition accuracy.

\section{Method}
\label{sec:methods}
\begin{figure*}[htbp]
	\begin{minipage}[htbp]{1.0\linewidth}
		\centering
		\centerline{\includegraphics[width=17cm]{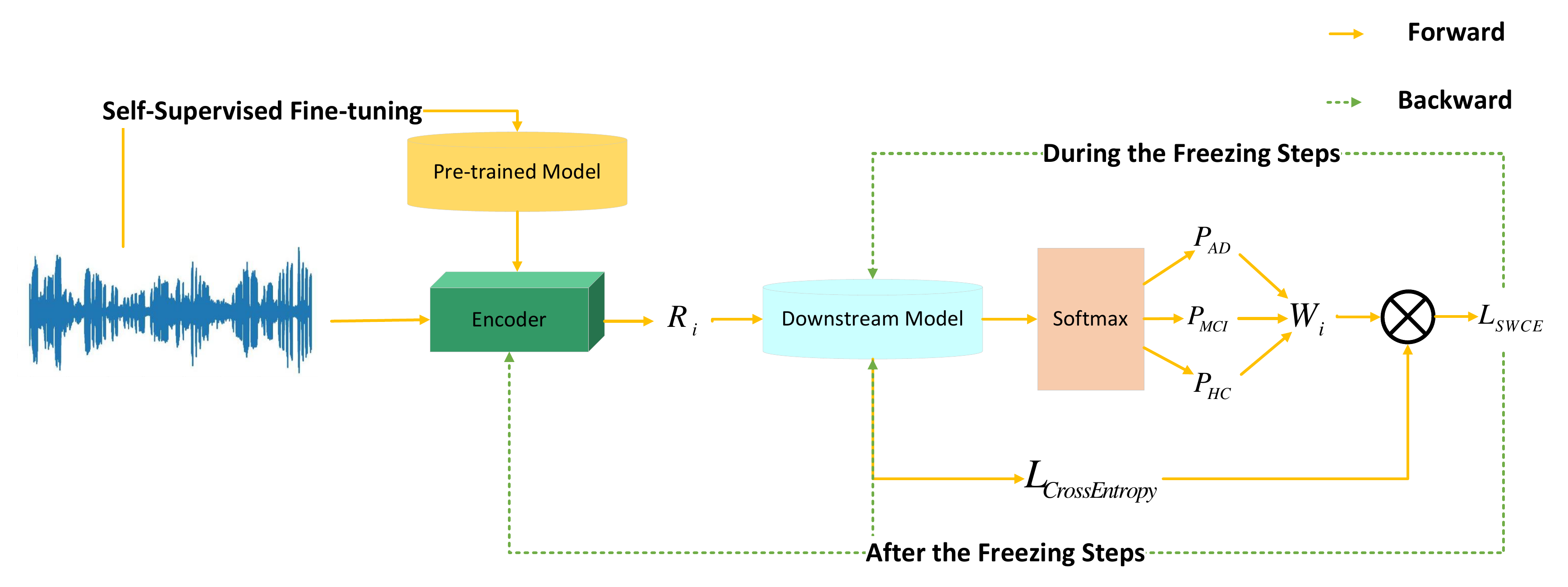}}
	\end{minipage}
	\caption{The architecture of the network with pre-trained model and downstream model.}
	\label{fig:res}
\end{figure*}

The key to solve the task of acoustic Alzheimer's disease recognition is obtaining effective representation features of original acoustic data and designing robust neural network models \cite{segovia2012comparative}. In this paper, we propose Wav2vec2 pre-trained model to extract acoustic feature from original audio and design Soft-Weighted CrossEntropy (SWCE) loss to improve the accuracy and generalization ability of our network. The model structure is presented on figure \ref{fig:res}, where $R_i$ represents the feature extracted by the pre-training model from the $i$-th segment of speech, and $W_i$ represents the weight of the $i$-th segment of speech calculated based on the soft weight cross-entropy loss.\\
Wav2vec2 has has achieved great success in the field of Automatic Speech Recognition (ASR), which indicates that the features extracted by Wav2vec2 completely preserve the vital information in the speech, which is also critical for the detection of Alzheimer’s disease. The Wav2vec2 consists of three stages. The first stage is a local encoder, consists of several blocks containing a temporal convolution followed by layer normalization and a GELU activation function. The second stage is a contextualized encoder with several layers of transformer with attention heads. Finally, a quantization module, takes the local encoder representations as input and turn the local encoder representation into logits. In our study, we extracted features from released Wav2vec2 framework and used pre-trained model XLSR-53 for Alzheimer's disease detection. We propose to optimize the pre-trained model XLSR-53 using trainable weights which are learned jointly with the downstream models. In our methods, the core of jointing training is parameters freezing and self-supervised fine-tuning. In the joint training of Wav2vec2 pre-trained model and downstream model, we update the parameters of downstream model after each epoch, and freeze the parameters of pre-trained model during the first N steps. Parameter freezing avoids changing the output of the pre-trained model before the downstream model is optimized to the extreme point, which not only speed up the convergence of joint training, but also improves the performance of Alzheimer's disease detection. As described in equation \ref{eq1}, \ref{eq2}, in the first N freezing steps, we only update the parameters of the downstream model, and then jointly train the pre-trained model and the downstream model. By setting the freezing steps to 0, 1000 and 2000 respectively, we compared the result of different freezing steps on the classification accuracy of joint training. 
\begin{equation}
Paras=[Para_{down}]\quad before\, N
\label{eq1}
\end{equation}
\begin{equation}
Paras=[Para_{pretrained}, Para_{down}]\quad after\, N
\label{eq2}
\end{equation}
In order to preserve more critical psychological information on the features extracted by Wav2vec2, we loaded the pre-trained model XLSR-53 based on Wav2vec2 framework, and applied the audios of NCMMSC2021 to self-supervised fine-tune the pre-trained model. The self-supervised fine-tuning of Wav2vec2 framework is similar to the masked language modeling used in BERT for Natural Language Processing (NLP). A certain proportion of contiguous time step $\rm{t}$ from the local encoder representations are masked during fine-tuning, and the model is trained to identifying the correct quantized representation $\rm{q_t}$ from $\rm{k+1}$ quantized candidate representations $\rm{\widetilde{q}}$. The objective of self-supervised fine-tuning is defined as equation \ref{eq3}:
\begin{equation}
L=\rm{-log}\dfrac{\rm{exp}(sim(c_t,q_t)/k)}{\sum_{\widetilde{q}\sim{Q_t}}\rm{exp}(sim(c_t,\widetilde{q})/k)}
\label{eq3}
\end{equation}
where $\rm{sim}(c_t,q_t)/k)$ represent the cosine similarity between the representations of context encoder and quantized latent speech representations.\\ 
The commonly used cross-entropy loss function only calculates the logarithmic sum of the prediction probability corresponding to the grand truth, without considering the difference between the probability of the label and the probability of other error categories. However, for a great performance classification model, we prefer the prediction probability of real tags to be greater than that of wrong tags. For example, for a data with three categories, if the real label is 0, we prefer the prediction probability of the model to be (0.8,0.1,0.1), rather than (0.4.0.3,0.3). In the Alzheimer's disease detection task, because there are great differences in speech between patients with Alzheimer's disease and healthy people, we hope that the results given by the model with great differences in the prediction probability between patients and the health. Therefore, we design the Soft-Weight Cross-Entropy (SWCE) loss function, as shown in formula \ref{eq5}. The calculation of soft-weights is shown in formula \ref{eq4}. The cross entropy loss value of each data is multiplied by a weight determined by the difference between the prediction probability of the real label and the probability of each wrong prediction. With the progress of training, the loss value decreases continuously, which increases the difference between the prediction probability of the real label and the probability of wrong prediction, so that the model can make more accurate judgment in the face of a new test data. Our research shows that the SWCE loss function of can significantly improve the performance of the model on the test set.
\begin{equation}
w_i=\rm{exp(-\dfrac{\sum_{j=0, j\neq{m}}^{N-1}p[m]-p[j]}{N})}
\label{eq4}
\end{equation}
\begin{equation}
L_{swce}=-\sum_{i=1}^{n}w_ip(x_i)\mathrm{log}q(x_i)
\label{eq5}
\end{equation}
where $\rm{p[m]}$ represents the probability of grand truth and $\rm{p[j]}$ is the probability of wrong prediction.
\section{Experiments}
\label{sec:result}
\subsection{Corpora and Tasks}
\label{ssec:data}
NCMMSC2021 Alzheimer's disease recognition challenge dataset is a Chinese Alzheimer's disease detection dataset, which is divided into two parts. The larger dataset includes 79 voice data of Alzheimer's disease patients, 93 Mild Cognitive Impairment (MCI) audios and 108 speech data of healthy controls (HC). The smaller data set contains 35 voice data of AD patients, 39 MCI speech segments and 45 HC audios. All the data in the NCMMSC2021 Alzheimer's disease recognition challenge dataset only has speech modal in Chinese. The average duration of each voice is about one minute, and the content is picture description and fluency naming. We take the smaller data set as the test set, randomly select 20\% speech data of the larger data set as the verification set, and the rest as the training set. All experiments use the Adam optimizer \cite{kingma2014adam}, the initial learning rate is 0.0001, and the batch size of all experiments is set to 8.
\subsection{Handcrafted Features and Downstream Models}
For the \textit{eGeMAPS} feature set has been widely used as a basic set of acoustic features based on their potential to detect physiological changes in voice production, we compare the performance of the representation extracted by Wav2vec2 pretrained model with some handcrafted features of \textit{eGeMAPS} on the same downstream model. We used the composite feature containing MFCC, F0 and constant-Q transform (CQT) as the input of the baseline \cite{sakar2019comparative} and applied a sliding window with a length of 250 ms on the audio files of the dataset with 50 ms overlap and extracted handcrafted features over such frames in order to encourage our model to discover the psychological information hidden in the frequency domain. The AD classification experiments were performed on two different downstream models, namely bidirectional Gated Recurrent Unit network (GRU) and Convolution Neural Network with self-attention mechanism (A-CNN). The GRU network includes two layers of GRU units as described in formula \ref{eq02}, where $f_j^L$ is the feature of the $L$ th frame and the $h$ is the hidden state of each direction. A-CNN is a convolution neural network with three convolution layers, two linear layers and one self-attention pooling layer.

\begin{equation}
\overrightarrow{h_k}=\mathrm{GRU}(f_j^{L},\overrightarrow{h_{k-1}})
\label{eq00}
\end{equation}
\begin{equation}
\overleftarrow{h_k}=\mathrm{GRU}(f_j^{L},\overleftarrow{h_{k-1}})
\label{eq01}
\end{equation}
\begin{equation}
Output_j=\mathrm{GRU}(\overleftarrow{h_k},\overrightarrow{h_k})
\label{eq02}
\end{equation}
\subsection{Results}
We have evaluate our network on NCMMSC2021 Alzheimer's disease recognition challenge dataset. The results of baselines for accuracy in the AD, MCI and HC classification task are summarised in Table 1. As indicated in boldface, the best performing classifier in test set was GRU, achieving 85.15\%. This indicates that when learning psychological information from low-dimensional acoustic features, the performance of GRU is significantly better than that of CNN.

\begin{table}[th]
	\caption{Detection accuracy (Acc) of Alzheimer's disease of GRU and A-CNN.}
	\label{tab1}
	\centering
	\setlength{\tabcolsep}{1.1mm}
	\begin{tabular}{ ccc@{}l  r }
		\toprule
		\multicolumn{1}{c}{\textbf{Model}} &
		\multicolumn{1}{c}{\textbf{Dev Acc($\%$)}} &
		\multicolumn{1}{c}{\textbf{Test Acc($\%$)}} \\
		\midrule
		GRU          & $80.21\pm{1}$   & $70.15\pm{1}$~~~       \\
		A-CNN        & $73.45\pm{1}$   & $64.83\pm{1}$~~~       \\
		\bottomrule
	\end{tabular}
\end{table}

We use the speech data of NCMMSC2021 datatset to self-supervised fine-tune the XLSR-53 pre-trained model based on Wav2vec2 framework and extract feature by pre-trained model XLSR after fine-tuning from the original speech. Compared with the baseline system, when the number of freezing steps is 0, the test accuracy of the system based on Wav2vec2 is reduced by 1.64\% on GRU but improved by 7.43\% on A-CNN, while when the number of freezing steps is 1000 and 2000, the features extracted by the pre-trained model have a positive gain on test accuracy for all downstream models. This indicates that in the Alzheimer's disease recognition task, the features extracted by the pre-trained model based on Wav2vec2 framework retain more psychological information than the handcrafted acoustic features. In addition, the recognition accuracy of the system when the freezing times are 1000 is better than that when the freezing times are 0 and 2000, which shows that in the current task, the freezing steps should not be too low, otherwise the parameters of the pre-trained model will be updated before the downstream model converges to the extreme point, so that the whole system is not easy to converge and also should not be too high to prevent the training of downstream model from finishing during freezing epochs, resulting in the pre-trained model parameters not being optimized.

\begin{table}[th]
	\caption{Performance of systems with different parameters freezing epochs.}
	\label{tab2}
	\centering
	\setlength{\tabcolsep}{1.1mm}
	\begin{tabular}{ cccc@{}l  r }
		\toprule
		\multicolumn{1}{c}{\textbf{Model}} &
		\multicolumn{1}{c}{\textbf{Epochs}} &
		\multicolumn{1}{c}{\textbf{Dev Acc($\%$)}} &
		\multicolumn{1}{c}{\textbf{Test Acc($\%$)}} \\
		\midrule
		XLSR-GRU    & $0$      	  & $76.32\pm{1}$            & $68.51\pm{1}$~~~       \\
		XLSR-GRU    & $1000$      & $\mathbf{89.34\pm{1}}$   & $\mathbf{76.48\pm{1}}$~~~       \\
		XLSR-GRU    & $2000$      & $87.52\pm{1}$            & $74.95\pm{1}$~~~       \\
		XLSR-A-CNN  & $0$         & $78.29\pm{1}$            & $72.17\pm{1}$~~~       \\
		XLSR-A-CNN  & $1000$      & $\mathbf{91.13\pm{1}}$   & $\mathbf{80.04\pm{1}}$~~~       \\
		XLSR-A-CNN  & $2000$      & $89.67\pm{1}$            & $76.93\pm{1}$~~~       \\
		\bottomrule
	\end{tabular}
\end{table}

As shown in Table 3, we applied the soft-weighted cross-entropy loss function to our model and compared it with the model that applied the cross-entropy loss function. The freezing epochs of all experiments were 1000. Although on the validation set, the recognition accuracy of the model with soft-weighted cross entropy loss has decreased, but it has increased by 5.03\% and 5.67\% on the test set. When using the cross-entropy loss function, our model has an accuracy difference of 12.86\% and 11.09\% on the validation set and test set, but this difference is reduced to 5.99\% and 1.79\% after using the soft-weighted cross-entropy loss. This indicates that the soft-weighted cross entropy loss can not only improve the recognition accuracy of model for test data that has never been used before, but also improve the generalization ability, so that the model has little difference between the verification set and the test set. This provides a method for some machine learning models to solve the problem of large difference between the performance of experimental datasets and practical applications.
\begin{table}[th]
	\caption{Comparison between the model with cross entropy loss ($L_{CE}$) and the model with soft-weight cross entropy loss ($L_{SWCE}$) in the detection accuracy of Alzheimer's disease.The parameter freezing epochs of the models are all 1000.}
	\label{tab3}
	\centering
	\setlength{\tabcolsep}{1.1mm}
	\begin{tabular}{ ccccc@{}l  r }
		\toprule
		\multicolumn{1}{c}{\textbf{Model}} &
		\multicolumn{1}{c}{\textbf{Loss}} &
		\multicolumn{1}{c}{\textbf{Dev Acc($\%$)}} &
		\multicolumn{1}{c}{\textbf{Test Acc($\%$)}} \\
		\midrule
		XLSR-GRU     &$L_{ce}$    & $89.34\pm{1}$   & $76.48\pm{1}$~~~       \\
		XLSR-GRU     &$L_{swce}$  & $87.50\pm{1}$   & $81.51\pm{1}$~~~       \\
		XLSR-A-CNN   &$L_{ce}$    & $91.13\pm{1}$   & $80.04\pm{1}$~~~       \\
		XLSR-A-CNN   &$L_{swce}$  & $87.50\pm{1}$   & $\mathbf{85.71\pm{1}}$~~~       \\
		\bottomrule
	\end{tabular}
\end{table} 

\section{Conclusion}
\label{sec:conclusion}
This paper presents a network for Alzheimer's disease detection from speech based on pre-trained model and downstream model. We use the speech data of National Conference on Man-Machine Speech Communication 2021 Alzheimer's disease recognition challenge to self-supervised fine-tune the pre-trained model XLSR-53 based on Wav2vec 2.0 framework, so that the features extracted by the pre-trained model could be more suitable for the task of Alzheimer's disease detection. We apply bidirectional Gated Recurrent Unit network and Convolution Neural Network with self-attention mechanism. The research indicates that the features extracted by pre-trained model after fine-tuning improve the recognition accuracy on the test set compared with the handcrafted acoustic features. In addition, we found that the parameters of the pre-trained model need to be frozen in the first N epochs in the process of joint training and the experimental results show that the system has the great performance when N is 1000. We also designed a Soft-Weighted Cross Entropy loss function, The purpose of this function is to increase the difference between the probability corresponding to the grand truth and the wrong prediction in the process of optimizing the network parameters. The experimental results indicate that on the National Conference on Man-Machine Speech Communication 2021 Alzheimer's disease recognition challenge dataset, the model using Soft-Weighted Cross Entropy loss improves the test accuracy by about 5\% compared with the model using cross entropy loss, and reduces the difference in accuracy between the verification set and the test set. The soft-weighted cross entropy loss provides a solution for some machine learning models to solve the problem of large difference in recognition accuracy between experimental datasets and application scenarios.



\bibliographystyle{IEEEbib}
\bibliography{Template}

\end{document}